\documentclass[a4paper,nodate]{sab}

\usepackage{graphicx} 
\usepackage{natbib} 
\bibpunct{(}{)}{;}{a}{}{,} 
\usepackage{url}
\usepackage[T1]{fontenc} 

\usepackage[utf8]{inputenc}
\usepackage[english]{babel}

\usepackage{amsfonts}
\usepackage{amsmath}

\usepackage{txfonts}

\usepackage{xcolor}
\usepackage[colorlinks,bookmarks]{hyperref}
\definecolor{linkblue}{rgb}{0,0,0.8}
\definecolor{linkgreen}{rgb}{0,0.5,0}
\hypersetup{linkcolor=linkblue, citecolor=linkgreen, urlcolor=linkblue}

\graphicspath{{./figs/}}

\def\d{{\rm d}}

\newcommand{\aperp}{a_{\perp }}

\newcommand{\apar}{a_{\parallel}}
\newcommand{\Hperp}{H_{\perp}}

\newcommand{\Hpar}{H_{\parallel}}

\renewcommand\ackname{Acknowledgements.} 
\renewcommand\andname{\&}
\renewcommand\keywordname{Keywords.} 
\renewcommand\figureptname{Figure}
\renewcommand\tableptname{Table}

\renewcommand\noteaddname{Note added to proofs.}
\renewcommand\andname{\&}

\idline{Boletim da Sociedade Astron\^omica Brasileira, {\bf 31}, no.~1, XX-XX}{1}

\begin{document} 
\title{Tensions in Cosmology}
\subtitle{Interpreting Them Through Inhomogeneous Models}

\author{Valerio Marra\inst{1}} 
\institute{Departamento de Física, Universidade Federal do Espírito Santo, 29075-910, Vitória, ES, Brazil\\\email{valerio.marra@me.com}}
\date{Received } 

\Abstract
{We review a subset of the current tensions affecting the standard $\Lambda$CDM cosmological model, emphasizing the role of chronic systematics and significance inflation in shaping their interpretation. As a unifying framework, we consider the spherically symmetric inhomogeneous $\Lambda$LTB model and use it as a set of ``glasses’’ through which to reinterpret the Hubble, dipole, and dark-energy tensions. Large-scale spatial gradients in this model introduce anisotropic expansion and position-dependent observables, allowing local estimates of $H_{0}$ to shift, dipolar signatures to arise, and an apparently evolving dark-energy equation of state to be mimicked without invoking genuinely dynamical dark energy. We discuss how these effects are constrained once the full supernova, CMB, and large-scale-structure data sets are included, and argue that it remains unclear whether any single $\Lambda$LTB configuration can simultaneously account for all major anomalies. More broadly, we highlight that cosmology currently lacks a widely accepted baseline model that is both theoretically well founded and capable of accommodating the Hubble and dark-energy tensions, leaving us without a true concordance framework for forecasting future surveys.}
{Revisamos um subconjunto das tensões atuais que afetam o modelo cosmológico padrão $\Lambda$CDM, destacando o papel dos erros sistemáticos crônicos e da inflação de significância na forma como essas tensões são interpretadas. Como estrutura unificadora, consideramos o modelo inhomogêneo esfericamente simétrico $\Lambda$LTB e o utilizamos como um conjunto de “lentes’’ para reinterpretar as tensões de Hubble, do dipolo e da energia escura. Gradientes espaciais de grande escala nesse modelo introduzem uma expansão anisotrópica e observáveis que dependem da posição, permitindo deslocar estimativas locais de $H_{0}$, gerar assinaturas dipolares e mimetizar uma equação de estado da energia escura aparentemente evolutiva sem recorrer a energia escura dinâmica. Discutimos como esses efeitos são restringidos quando os conjuntos completos de dados de supernovas, CMB e estrutura em larga escala são incluídos, e argumentamos que ainda não está claro se uma única configuração $\Lambda$LTB pode explicar simultaneamente todas as principais anomalias. De forma mais ampla, destacamos que a cosmologia atualmente carece de um modelo de referência amplamente aceito, teoricamente bem fundamentado e capaz de acomodar as tensões de Hubble e da energia escura, deixando-nos sem um verdadeiro modelo de concordância para previsões em missões futuras.}



\maketitle 

\section{Introduction}

By assuming large-scale statistical isotropy and homogeneity, encoded in the Friedmann-Lemaître-Robertson–Walker (FLRW) metric, cosmology has made remarkable progress in describing the Universe. Within this framework, we have obtained a quantitative account of cosmic evolution from the earliest times to the present, across all observable scales. This success has established the $\Lambda$ cold-dark-matter (CDM) model as the standard model of cosmology.

In the $\Lambda$CDM model the Universe is dominated by a cold, non-baryonic component called dark matter and by a dark energy component responsible for the present accelerated expansion, often modeled as a cosmological constant $\Lambda$, that is, a tiny vacuum energy. Dark matter drives the growth of structure and the formation of galaxies, while dark energy governs the late-time accelerated expansion of the Universe. Together they constitute about 95\% of the cosmic energy budget, yet a satisfactory theoretical understanding of this ``dark sector’’ is still lacking.

The appeal of $\Lambda$CDM lies in its simplicity: assuming General Relativity and small perturbations around a spatially homogeneous and isotropic background, it fits tens of thousands of data points, probing a wide range of scales in space and time, with only a handful of parameters. However, over the last decade several independent observations have been found to deviate from the model’s predictions at the $2$–$5\sigma$ level. In what follows we will focus on a subset of these tensions and refer the reader to \citet{Perivolaropoulos:2021jda,CosmoVerseNetwork:2025alb} for a comprehensive discussion of the more and less established anomalies within the standard model of cosmology.

These tensions may be signaling physics beyond $\Lambda$CDM. A key requirement for any alternative cosmological model is its ability to address not just a single anomaly, but several tensions simultaneously. Individual discrepancies are often later traced back to previously unrecognized systematics or analysis choices. In contrast, a beyond-$\Lambda$CDM model that naturally alleviates multiple, independent tensions simultaneously would both lend credence to the physical reality of those tensions and effectively enhance their combined statistical significance, notwithstanding the systematic uncertainties that are intrinsic to cosmological observations.

\begin{figure*}
\centering 
\includegraphics[trim={0 0 0 0}, clip, width= \linewidth]{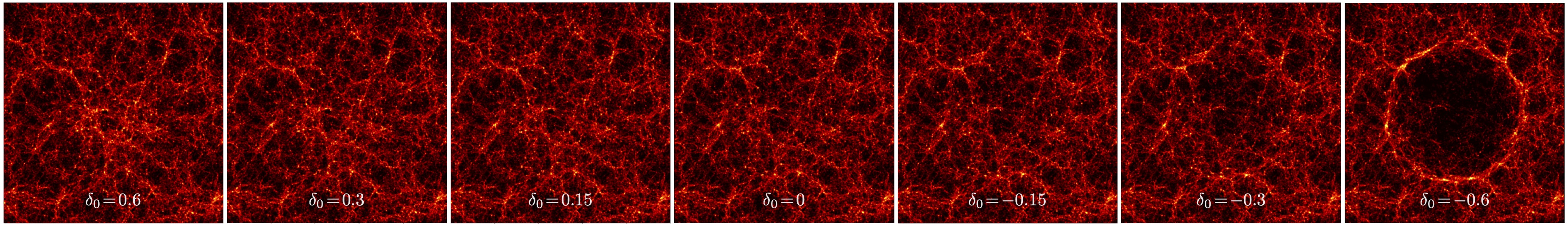}
\caption{Growth of structures in an inhomogeneous background. From \href{https://valerio-marra.github.io/BEHOMO-project/}{valerio-marra.github.io/BEHOMO-project}.}
\label{fig:banner}
\end{figure*}

\section{Chronic systematics}

In cosmology and astronomy, it is inherently difficult to deal with systematic errors that can bias an experiment. While instrumental effects can often be characterized and kept under reasonable control, systematics associated with cosmological sources at large distances are much harder to quantify: modeling choices and assumptions play a central role in interpreting light that was emitted billions of years ago. Systematic biases have traditionally been searched for by testing specific, preconceived possibilities, but this strategy is intrinsically ill-suited to account for ``unknown unknowns.''

For this reason, astronomers and cosmologists are always skeptical of interpreting tensions in the data as evidence for new physics. The natural mindset is that one has uncovered a new, interesting systematic effect rather than a fundamental discovery. Consequently, the acceptance of a tension as genuine requires both a high level of scrutiny of the modeling and experimental pipeline and a correspondingly higher statistical threshold.

\section{Significance inflation and the look-elsewhere effect}

A further aspect that complicates the interpretation of cosmological tensions is significance inflation,\footnote{\url{https://xkcd.com/882/}} also known as the look-elsewhere effect. When many data cuts, statistics, parameter combinations, or models are examined, the probability of finding at least one apparently significant deviation from $\Lambda$CDM purely by chance increases. If one then reports only the most anomalous result with its local significance, as if it had been predicted in advance, the quoted significance is effectively inflated.

Formally, one distinguishes between local and global significance. The local significance quantifies how unlikely a given fluctuation is at a specific, pre-defined location in parameter or data space. The global significance, instead, asks how likely it is to obtain a fluctuation at least as extreme as the observed one anywhere in the full space that was actually searched. Because cosmological analyses routinely explore many redshift bins, angular scales, summary statistics, and model variants, the global significance is always lower than the local one, sometimes substantially so.

In practice, correctly accounting for the look-elsewhere effect is challenging. It requires a careful characterization of how many ``effective trials'' have been performed, often via extensive simulations. Compounding this difficulty is a sociological effect: there is no ``Journal of Failed Tests'' systematically publishing null results and non-detections. As a consequence, the literature is naturally biased toward successful or anomalous findings, so that our main journals effectively behave as ``journals of outliers.'' This publication bias, combined with the look-elsewhere effect, can lead to an overabundance of reported $2$–$3\sigma$ anomalies, many of which are expected to arise even in a perfectly correct model. Recognizing and quantifying significance inflation is therefore essential before promoting a given tension to evidence for new physics.

\section{The $\Lambda$LTB model}

\begin{figure*}
\centering 
\includegraphics[trim={0 0 0 0}, clip, width= .9 \linewidth]{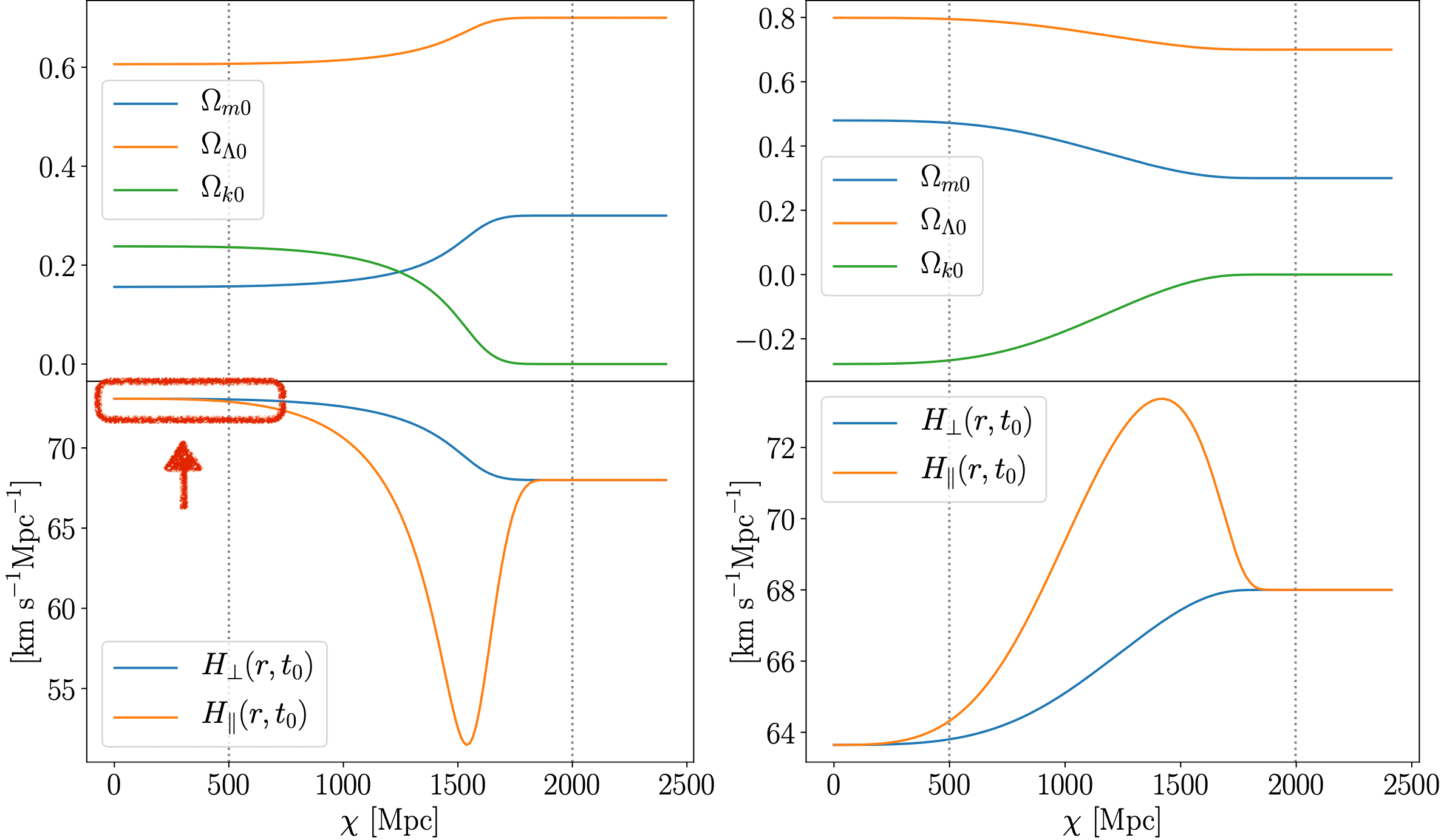}
\caption{Density parameters (top) and Hubble rates (bottom) for an illustrative underdensity (left) and overdensity (right).}
\label{fig:example-t0}
\end{figure*}

For concreteness, we adopt a specific extension of the standard model as a common framework to discuss the tensions, namely the spherically symmetric $\Lambda$LTB model.\footnote{\url{https://en.wikipedia.org/wiki/Spherical_cow}}
We use the Lemaître-Tolman-Bondi (LTB) metric to describe the impact of large-scale spatial gradients, whose generic consequence is an anisotropic expansion encoded in the background shear. This shear induces both large-scale bulk flows and modifies the growth rate of structures, as illustrated in Figure~\ref{fig:banner}.

This large-scale inhomogeneity can be viewed as a particular realization of primordial non-Gaussianity, characterized by non-standard amplitudes and phases of long-wavelength modes. It generates bulk flows and coherent perturbations in the energy content of the Universe on very large scales. As a result, large-scale homogeneity and isotropy are effectively broken by the phases of these additional modes, and cosmological observables acquire an explicit dependence on the position of the observer.

The $\Lambda$LTB model is defined by a metric and a Friedmann-like equation which, with suitable definitions, resemble their FLRW counterparts, but with an explicit radial dependence in the spatial curvature, scale factors, and matter density:
\begin{align} 
&\d s^2 = -\d t^2 + \frac{\apar^2(t,r)}{1-k(r)r^2}\d r^2 + \aperp^2(t,r)r^2 \, \d\Omega^2 \,, \label{metric} \\
&\Hperp^2(t,r) = {8 \pi G \over 3}   \rho_m^{\rm e}(t,r)  +  {8 \pi G \over 3} \rho_\Lambda-  { k(r) \over \aperp^2(t,r)} \,,  \label{fried}
\end{align}
where we refer to \citet{Marra:2022ixf} for a comprehensive review of $\Lambda$LTB models.

As already mentioned, the general effect of spatial gradients is to introduce anisotropic expansion, here manifested by the presence of two distinct scale factors: the longitudinal ($\apar$) and perpendicular ($\aperp$) scale factors. They define two different Hubble rates:
\begin{align}
\Hperp(t,r) \equiv \frac{\dot a_\perp}{\aperp} = \frac{\dot Y}{Y} \,,
\qquad
\Hpar(t,r) \equiv \frac{\dot a_{\|}}{\apar} = \frac{\dot Y'}{Y'} \,,
\end{align}
where a dot denotes differentiation with respect to the coordinate time $t$. This has important implications when confronting these models with observations, which can be used to constrain the background shear,
\begin{align}
\Sigma(t,r) = \frac{2}{3} \bigl[ \Hpar(t,r) - \Hperp(t,r) \bigr] \,.
\end{align}
Figure~\ref{fig:example-t0} shows the density parameters and Hubble rates for an illustrative underdensity (left) and overdensity (right).

\section{The Hubble Tension}

Local distance-ladder calibration, primarily based on Cepheids and Type Ia supernovae, yields $H_{0}=73.04 \pm 1.04 \,\mathrm{km\,s^{-1}Mpc^{-1}}$ \citep{Riess:2021jrx}, while the CMB inference within $\Lambda$CDM gives $H_{0}=67.36 \pm 0.54 \,\mathrm{km\,s^{-1}Mpc^{-1}}$ \citep{Planck:2018vyg}. The resulting discrepancy exceeds $5\sigma$.

This tension has become more pronounced over the past decade, and, although a recent meta-analysis \citep{H0DN:2025lyy} reports \(H_{0}=73.50 \pm 0.81 \,\mathrm{km\,s^{-1}Mpc^{-1}}\) with a significance exceeding \(7\sigma\) relative to the combined Planck+SPT+ACT constraint \(H_{0}=67.24 \pm 0.35 \,\mathrm{km\,s^{-1}Mpc^{-1}}\) \citep{SPT-3G:2025bzu}, and finds no evidence for uncontrolled systematics, confirmation from an independent probe remains highly desirable.

Bright sirens offer a promising avenue, as they are subject to a distinct and more limited set of systematics. However, the results of \citet{Menote:2025dew}, based on the events forecasted in \texttt{CosmoDC2\_BCO}\footnote{\url{https:/github.com/LSSTDESC/CosmoDC2_BCO}} \citep{Menote:2025zmn}, indicate that current-generation detectors will not have sufficient constraining power to decisively confirm or refute the tension. A clear resolution would occur only if the measured expansion rate ends up  below the Planck value or  above the local distance-ladder determination.

\begin{figure}
\centering 
\includegraphics[trim={0 0 0 0}, clip, width= .7 \linewidth]{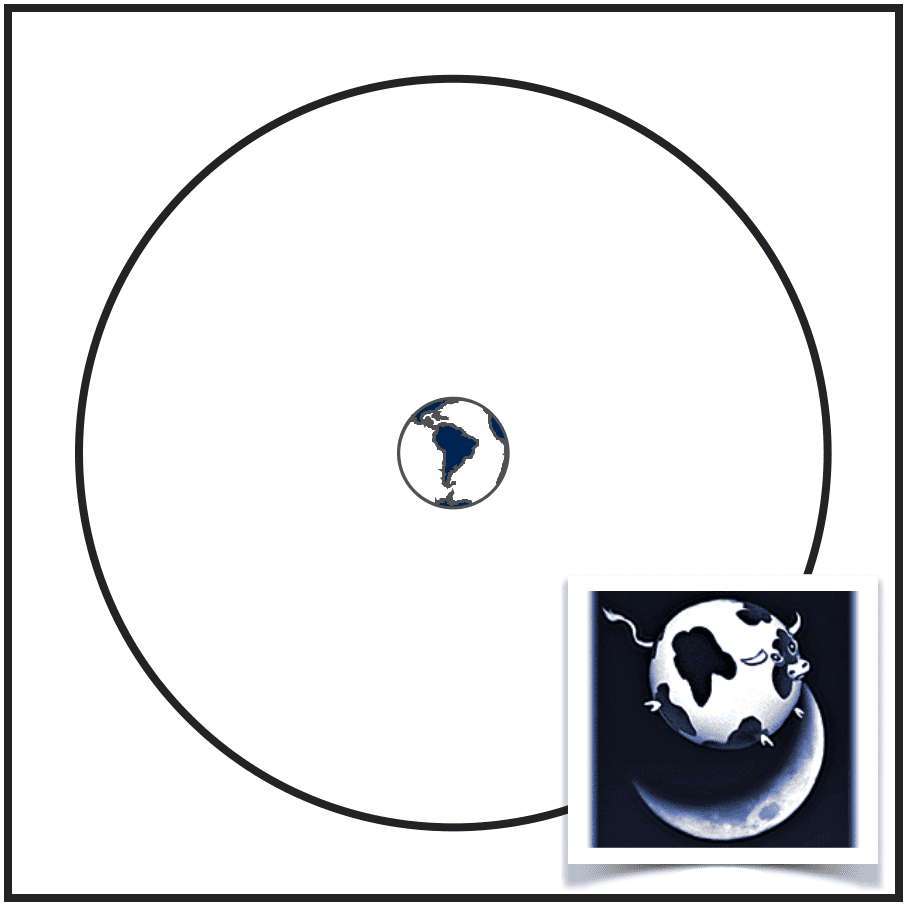}
\caption{An observer near the center of an underdensity measures an enhanced local expansion rate, as illustrated in Figure~\ref{fig:example-t0}.}
\label{fig:h0}
\end{figure}

An observer located at the center of an underdense $\Lambda$LTB region (Figure~\ref{fig:h0}) could, in principle, account for part of the Hubble tension, as illustrated in Figure~\ref{fig:example-t0} (bottom left panel). However, the standard spectrum of perturbations inferred from the CMB can generate at most a $0.5\mathrm{-}0.9\,\mathrm{ km\,s^{-1}Mpc^{-1}}$ cosmic-variance contribution to the local Hubble constant \citep{Marra:2013rba,Camarena:2018nbr}, which falls far short of explaining the observed discrepancy.

\begin{figure}
\centering 
\includegraphics[trim={0 0 0 0}, clip, width=  \linewidth]{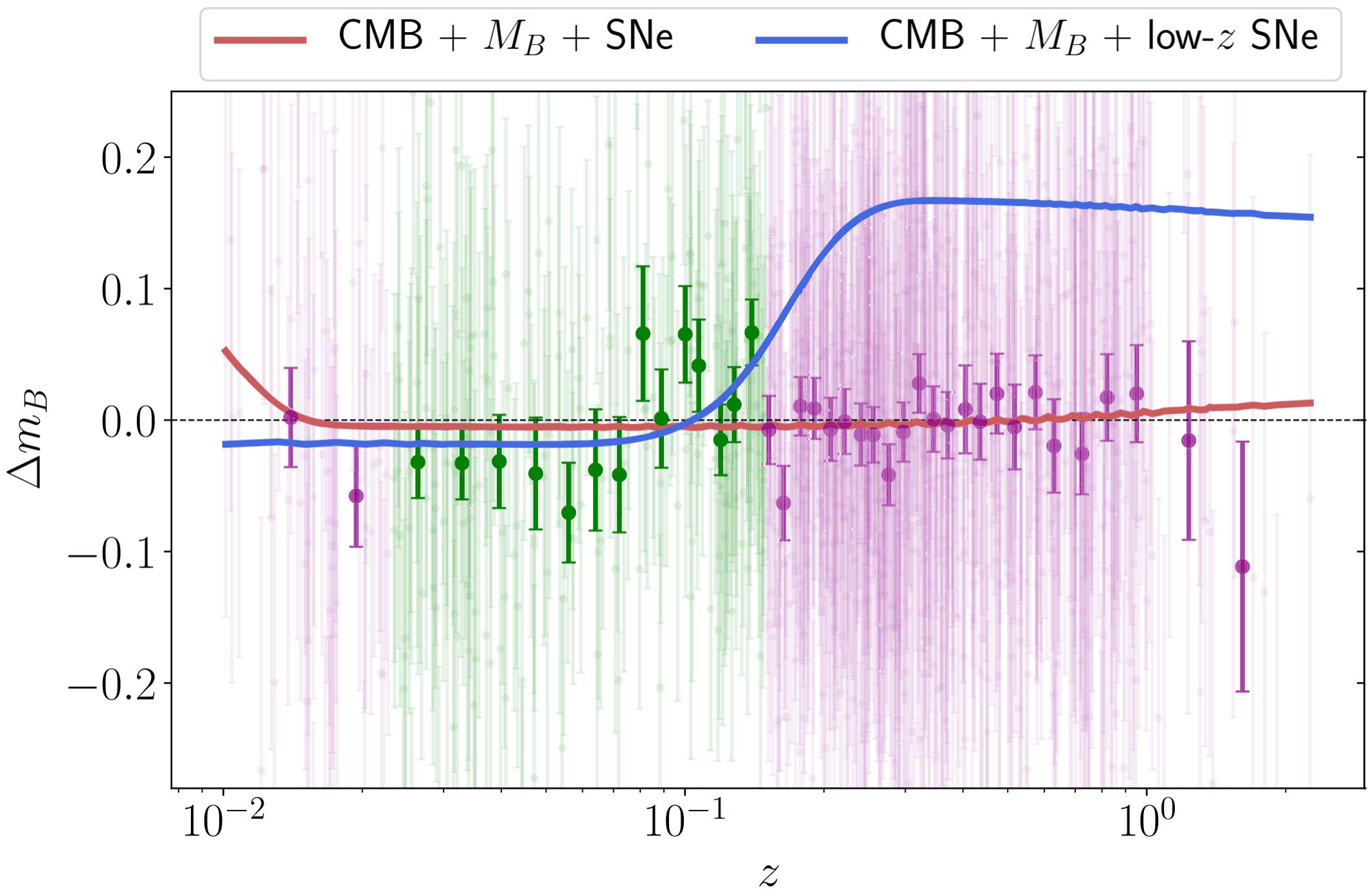}
\caption{$\Lambda$LTB model fitted to CMB and calibrated supernova data. The green points indicate the subset used to measure the local $H_{0}$. The full supernova sample does not permit the luminosity jump required to explain away the $H_{0}$ tension. From \citet{Camarena:2022iae}.}
\label{fig:avoid}
\end{figure}

From a purely phenomenological perspective, one may then attempt to fit a generic $\Lambda$LTB model directly to the data to determine whether the tension can be absorbed by a suitably chosen radial profile. This approach succeeds only when using local supernova samples, but fails once the full dataset is included, as shown in Figure~\ref{fig:avoid}: the supernova luminosity–distance relation does not permit the abrupt change required to reconcile a significantly higher local expansion rate with the global expansion history.

Interestingly, the best-fit model from \citet{Camarena:2022iae} suggests a shallow underdensity of about 5\%, consistent with the constrained local-universe simulations of \citet{McAlpine:2022art}. An ellipsoidal description of the local environment is discussed in \citet{Giani:2023aor}.

\section{Dipole Anomalies}

\begin{figure}
\centering 
\includegraphics[trim={0 0 0 0}, clip, width=  \linewidth]{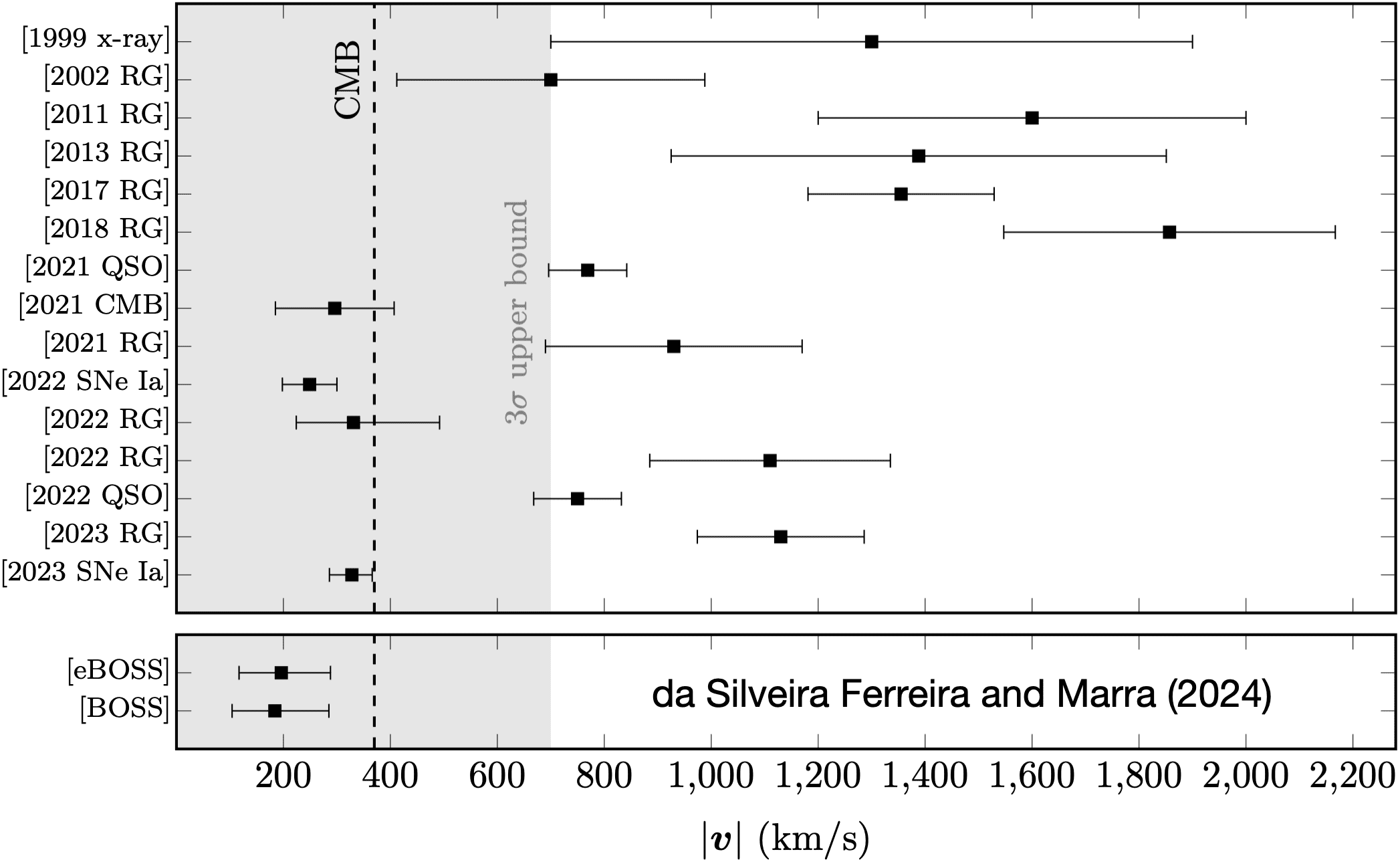}
\caption{
Velocity estimates inferred from dipoles shown in chronological order with $1\sigma$ uncertainties.
Top: previous measurements based on X-ray sources, radio galaxies (RG), quasars (QSO), CMB non-diagonal correlations, and SNe Ia.
Bottom: results from \citet{daSilveiraFerreira:2024ddn}, combining multiple tracers over $0.4<z<2.2$.
Dashed line: velocity implied by the CMB dipole under the peculiar-velocity interpretation. From \citet{daSilveiraFerreira:2024ddn}.}
\label{fig:dipole}
\end{figure}

\begin{figure}
\centering 
\includegraphics[trim={0 0 0 0}, clip, width= .7 \linewidth]{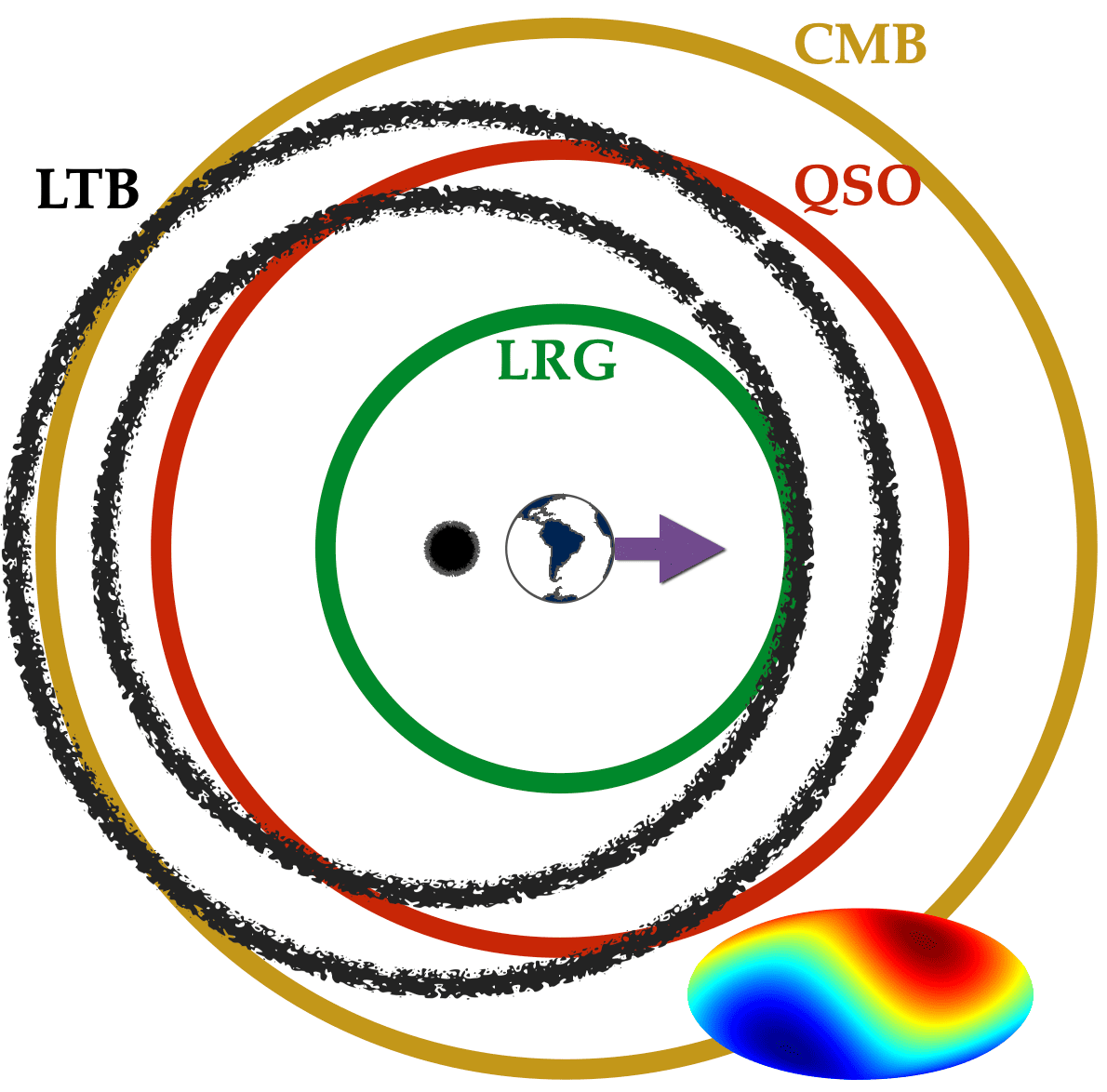}
\caption{Off-center $\Lambda$LTB models can reconcile dipole velocities from different tracers.}
\label{fig:dipole-ltb}
\end{figure}

According to the \citet{1984MNRAS.206..377E} test, velocity estimates of our motion with respect to the large-scale FLRW background should be mutually consistent and, in particular, should agree with the exquisitely measured CMB dipole. Over the past 25 years, however, several determinations—especially those based on radio galaxies \citep{Wagenveld:2023kvi} and QSO number counts \citep{Secrest:2022uvx}—have reported dipole amplitudes exceeding the expectation from the CMB dipole (see Figure~\ref{fig:dipole} for a summary), suggesting a possible breakdown of large-scale homogeneity and isotropy.
Figure~\ref{fig:dipole-ltb} illustrates how a $\Lambda$LTB model with an off-center observer could provide the flexibility required to accommodate the different velocities inferred from distinct tracers. Such as setup has been recentely investigated by \citet{Cai:2022dov}.

However, \citet{daSilveiraFerreira:2024ddn}, using a tomographic redshift dipole estimator applied to BOSS and eBOSS galaxies and quasars, and calibrated with realistic mocks, found that the eBOSS sample yields a velocity field fully consistent with the CMB dipole, in clear contrast with earlier number-count results. This reinforces the cosmological principle across gigaparsec scales. By comparison, the BOSS data revealed unmodeled systematics, with one sample not allowing a stable fit and another producing a dipole direction incompatible with the CMB. Understanding these systematics and the tension with number-count analyses remains essential for robustly testing large-scale homogeneity and isotropy. Upcoming DESI measurements may help clarify this issue.

\section{Dark-Energy Tension}

DESI BAO in combination with CMB observations \citep{DESI:2025zgx} favor evolving dark energy at more than 3$\sigma$, with a transition to $w<-1$ for $z\gtrsim 0.5$. Such a phantom crossing of $w=-1$ is theoretically problematic in simple scalar-field models and may instead signal a nontrivial interaction between dark matter and dark energy \citep{Petri:2025swg}. It is therefore natural to ask whether this tension reflects genuinely dynamical dark energy, or whether part of it could be attributed to large-scale inhomogeneities.

Within the $\Lambda$LTB framework, an observer located near the center of a spherically symmetric inhomogeneity interprets spatial gradients in the expansion rate as temporal variations, because light-cone observables mix time and radius according to
\begin{align}
\frac{\d}{\d t} \approx \frac{\partial}{\partial t} - \frac{\partial}{\partial r} \,.
\end{align}
An apparent evolution of the effective equation of state, including deviations from $w=-1$, can therefore arise from radial gradients in the background rather than from a genuinely dynamical dark energy component. When confronted with data that suggest $w(z)\neq -1$, one should thus carefully disentangle dynamical effects from geometric ones induced by large-scale inhomogeneities.

\begin{figure}
\centering 
\includegraphics[trim={0 0 0 0}, clip, width= \linewidth]{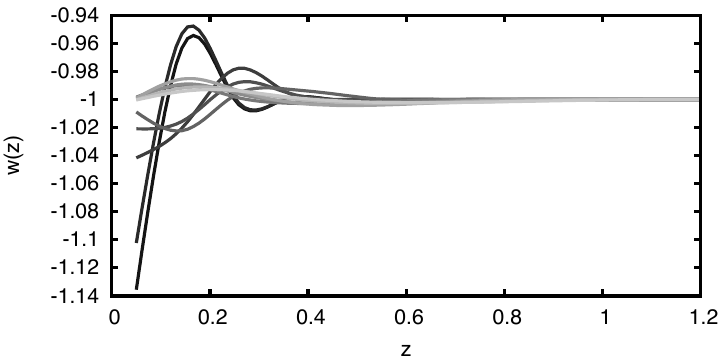}
\caption{Examples of apparent evolution of the dark-energy equation of state that an FLRW observer would deduce from observations in a flat $\Lambda$CDM universe endowed with a local inhomogeneity coming from a post-inflationary gaussian density field. From \citet{Valkenburg:2013qwa}.}
\label{fig:wnoise}
\end{figure}

This point was emphasized clearly by \citet{Valkenburg:2013qwa}. Figure~\ref{fig:wnoise} illustrates how naturally a background inhomogeneity can generate an apparent equation of state that crosses \(w=-1\). More recently, \citet{Camarena:2025upt} showed that DESI data may already contain first indications of the anisotropic expansion rate induced by large-scale spatial gradients.

\section{Conclusions}

The collective interpretation of current cosmological tensions demands considerable caution. Significance inflation, together with a publication bias toward successful or anomalous results, complicates a robust statistical assessment. Many reported discrepancies arise from complex analyses in which calibration uncertainties, photometric systematics, or modeling choices can propagate into cosmological inferences in subtle ways.

At the same time, the persistence and multi-probe nature of some anomalies, most notably the Hubble and dark-energy tensions, make it increasingly difficult to regard them as mere statistical fluctuations. A convincing physical explanation would ideally connect several of these discrepancies within a unified framework, whether through modified gravity, additional fields, nonstandard early-Universe physics, or departures from large-scale homogeneity and isotropy.

In this spirit, we used the $\Lambda$LTB inhomogeneous model as a set of ``glasses’’ through which to reinterpret several anomalies. Spatial gradients can mimic an evolving equation of state, shift local estimates of $H_{0}$, and generate dipolar signatures. However, when confronted with the full suite of observations, these models are tightly constrained, and it remains unclear whether any single $\Lambda$LTB configuration—with an observer placed at a specific radius—can simultaneously alleviate multiple tensions.

More broadly, we are in the unsatisfactory situation that there is no widely accepted cosmological model that is both theoretically well grounded and able to accommodate the Hubble and dark-energy tensions. The $\Lambda$CDM paradigm, despite its conceptual puzzles in the dark sector, still provides an excellent global fit to most high-precision data, yet it does not naturally resolve these anomalies. Proposed extensions often ease one discrepancy at the cost of new fine-tunings, ad hoc ingredients, or tensions elsewhere. As a result, the field currently lacks a successor to $\Lambda$CDM that can serve as a theoretically compelling and empirically superior concordance model.
Simply put, when performing forecasts for future missions, we lack a true concordance model to serve as a reliable baseline.

Whether the present tensions ultimately trace back to subtle systematics or constitute the first clear evidence for new physics will be tested decisively by upcoming surveys. DESI, Euclid, Rubin, J-PAS, Roman, and next-generation gravitational-wave detectors will deliver data with unprecedented volume and quality. Progress will rely not only on more precise measurements but also on joint multi-probe analyses, cross-calibration between experiments, and realistic end-to-end simulations. These efforts will be essential to resolve the current discrepancies, to clarify the role of large-scale inhomogeneities, and, ultimately, to refine or replace the standard cosmological model that underpins our description of the Universe.

\begin{acknowledgements}
VM acknowledges partial support from CNPq (Brazil) and FAPES (Brazil).
\end{acknowledgements} 

\bibliographystyle{aa-style}
\bibliography{refs}

\end{document}